\documentclass[usenatbib]{mn2e}
\usepackage{apjfonts,amsfonts,amsmath,amssymb,bm,ctable,verbatim}

\newcommand{\acknowledgments}[1]{\begin{small}\section*{Acknowledgments}\end{small}{\noindent #1}\vspace{5pt}}

\newif\ifshowcomments
\showcommentstrue

\title[The mass budget for IMBHs in star clusters]{The mass budget for intermediate-mass black holes in dense star clusters}

\author[Shi et al.]{
Yanlong Shi,$^{1}$\thanks{yanlong@caltech.edu}
Michael Y. Grudi\'{c},$^{1,2}$
Philip F. Hopkins $^{1}$
\\
$^{1}$TAPIR, Mailcode 350-17, California Institute of Technology, Pasadena, CA 91125, USA\\
$^{2}$Department of Physics and Astronomy and CIERA, Northwestern University, 1800 Sherman Ave, Evanston, IL 60201, USA}

\date{}
\begin{document}
\maketitle

\begin{abstract}
Intermediate-mass black holes (IMBHs) could form via runaway merging of massive stars in a young massive star cluster (YMC). We combine a suite of numerical simulations of YMC formation with a semi-analytic model for dynamical friction and merging of massive stars and evolution of a central quasi-star, to predict how final quasi-star and relic IMBH masses scale with cluster properties (and compare with observations). The simulations argue that inner YMC density profiles at formation are steep (approaching isothermal), producing some efficient merging even in clusters with relatively low effective densities, unlike models which assume flat central profiles resembling those of globular clusters (GCs) {\em after} central relaxation. Our results can be approximated by simple analytic scalings, with $M_{\rm IMBH} \propto v_{\rm cl}^{3/2}$ where $v_{\rm cl}^{2} = G\,M_{\rm cl}/r_{\rm h}$ is the circular velocity in terms of initial cluster mass $M_{\rm cl}$ and half-mass radius $r_{\rm h}$. While this suggests IMBH formation is {\em possible} even in typical clusters, we show that predicted IMBH masses for these systems are small, $\sim 100-1000\,M_{\odot}$ or $\sim 0.0003\,M_{\rm cl}$, below even the most conservative observational upper limits in all known cases. The IMBH mass could reach $\gtrsim 10^{4}\,M_{\odot}$ in the centers nuclear star clusters, ultra-compact dwarfs, or compact ellipticals, but in all these cases the prediction remains far below the present observed supermassive BH masses in these systems.
\end{abstract}


\section{Introduction}
\label{sec:intro}

Intermediate massive black holes (IMBHs), which typically weigh $10^2\textrm{--}10^5~M_{\odot}$, are believed to be the missing link between stellar mass black holes and super massive black holes (SMBHs). These objects, if they exist, are expected to play an important role in multiple astrophysical processes, e.g., affecting the evolution of globular star clusters and powering off-nuclear ultraluminous X-ray sources (ULXs). More importantly, they are potentially the progenitors of SMBHs which are known to live in most galaxies \citep{gebhardt2000smbh,ferrarese2000smbh,volonteri_formation_2010,mezcua_observational_2017,koliopanos_intermediate_2018}. Observations of ULXs and stellar kinematics argued that there may be some evidence for such objects in galaxies \cite[e.g.,][]{farrell_intermediate-mass_2009, kaaret_2017_ULX_review}, and globular star clusters \cite[e.g.,][see more in \S \ref{sec:discussion}]{portegies_zwart_formation_2004}. But these claims remain controversial.
  
Theoretically, several different IMBH formation channels have been proposed. Major ideas include: direct collapse of hyper-mass quasi-stars in isolation \cite[e.g.,][]{volonteri_quasi-stars_2010,schleicher_massive_2013}, runaway hyper-Eddington accretion onto stellar mass black holes \cite[e.g.,][]{ryu2016intermediate}, and runaway mergers in globular (star) clusters \cite[GCs, e.g.,][]{portegies_zwart_runaway_2002,gurkan_formation_2004}.  All these mechanisms have challenges. For the direct collapse channel, the fragmentation of molecular clouds may not form quasi-stars instantly, but star clusters \cite[]{moran2018effects}. For the hyper-Eddington channel, one important question is whether such high-efficiency accretion is sustainable or even possible. Finally, for the runaway merger channel, gravitational recoil due to merging stellar-mass BHs will likely ``kick'' the IMBH to a high velocity \cite[$\sim 1000~{\rm km/s}$, e.g.,][]{holley-bockelmann_gravitational_2008}, sufficient to  make it escape the star cluster. Only relatively massive IMBHs ($\gtrsim 10^3~M_\odot$) could remain in the galactic field even after the star cluster dissolves and survive such a merger without too-large a ``kick'' \cite[]{fragione_gravitational_2018}, which means the IMBH must be that massive before stellar-mass BH mergers occur. To solve this problem, the runaway merging process must be rapid enough such that massive stars merge together \emph{before} they evolve off the main sequence and become black holes individually.
  
\begin{figure*}
    \includegraphics[width=\textwidth]{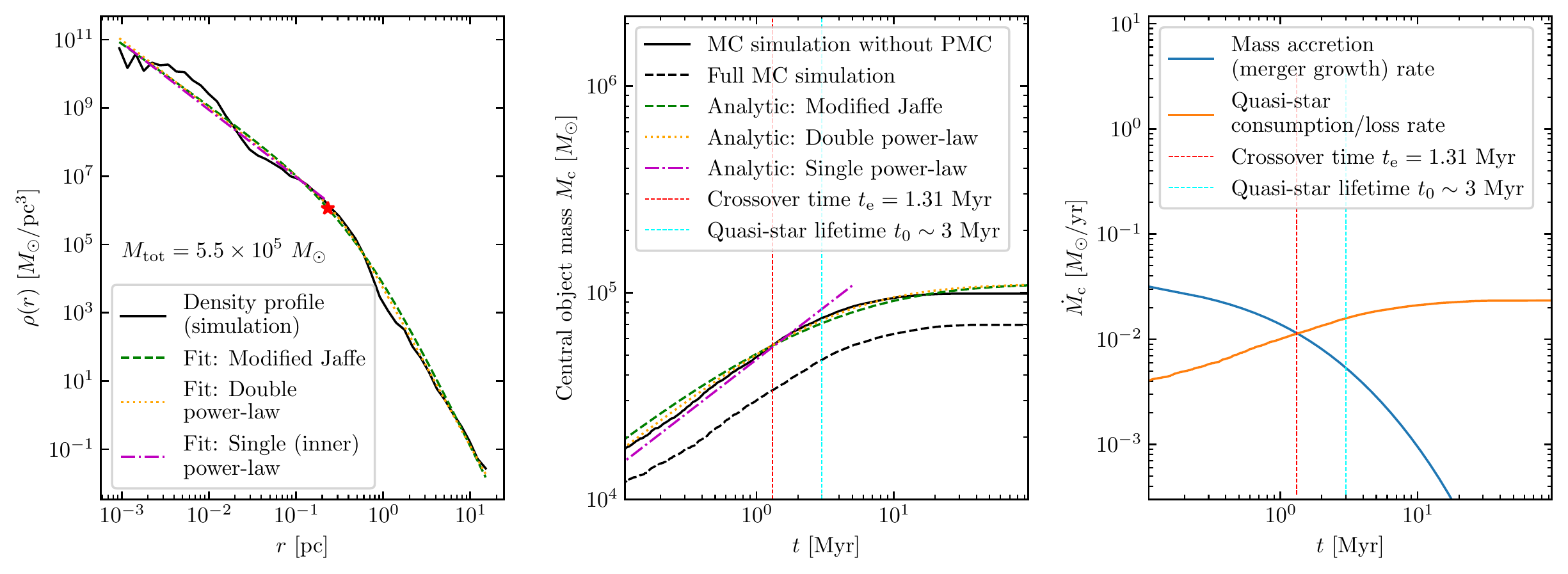}
    \vspace{-0.7cm}
    \caption{\textbf{Left}: Radial density profile $\rho(r)$ for a young, massive,  star cluster (stellar mass $M_{\rm tot}=5.5 \times10^{5}\,M_{\odot}$) in the studied suite of star-cluster formation simulations, taken just after the peak episode of star formation. We compare three analytic fits to the profile: ``modified Jaffe'' (Eq.~\ref{equ:jaffe}), ``double power-law'' (Eq.~\ref{equ:doublepower}), and a single power-law fit only to the ``inner'' radii within the half-mass radius $r_{\rm h}$ (marked by a red star). 
    \textbf{Middle}: Calculated mass growth of the central massive object, owing to mergers of massive stars. {\it Full MC simulation}: calculating the inspiral of massive stars in the cluster sampling from the IMF and following each orbit as described in (\S~\ref{sec:methods}), identifying them as merged when they contact the quasi-stellar radius. {\it Crossover time}: time when the accretion timescale $\dot{M}_{\rm acc}/M_{\rm c}$ becomes $>3\,$Myr, an estimate of when mass-loss from the quasi-star might outpace accretion). {\it MC simulation without PMC}: MC simulation ignoring the ``point mass correction'' (``without PMC'', also refer \S~\ref{sec:method:sink.merge}) -- i.e. ignoring the effect of the quasi-star itself and finite mass $N$-body effects in the center of the star cluster on stellar dynamics (instead assuming the density profile is simply the smooth/continuous extrapolation of the continuous $\rho(r)$. This produces systematically higher $M_{\rm c}$ as the PMC causes inspiralling stars to stall, but the effect is relatively small (tens of percents)  {\it Analytic:} Closed-form, approximate solutions for $M_{\rm c}(t)$, using the analytic fits ({\em left}). These agree well with the MC without PMC model, so the PMC is the dominant correction.
    \textbf{Right:} Mass accretion rate $\dot{M}_{\rm c}$ history of the central massive object. We compare a fuel consumption+stellar mass loss rate for the quasi-star given by a toy model for Eddington-limited growth, with $\dot{M}_{\rm c,\,consumption+loss} \sim -M_{\rm c}/{\rm 3\,Myr}$. At early times, growth rates are much larger than this loss rate, while at later times, accretion rates drop rapidly. As a result, the exact assumption about when to ``truncate'' accretion rates and how to model quasi-star mass-loss make relatively little difference to our predictions for $M_{\rm c}$ (though they are important for models which attempt to predict the {\em relic} IMBH mass, given some $M_{\rm c}$). 
    \label{fig:fit-evolution}}
\end{figure*}
  
Previous studies of runaway mergers in star clusters have provided us with a possible scenario: relatively massive main-sequence stars sink to the cluster's center due to dynamic friction from the background, the stars then merge into a supermassive quasi-star which then self-collapses to an IMBH after $\sim$3 Myr. In \cite{portegies_zwart_runaway_2002}, the authors showed $N$-body simulations of the process, and found that star clusters with initial half-mass relaxation time scale $t_{\rm rlx}\lesssim 25~{\rm Myr}$ can form IMBHs. 
More precise simulations in \cite{gurkan_formation_2004} drew a similar conclusion and predicted that the quasi-star's mass could account for $\sim0.1\%$ of the total cluster mass. The studies focused on cluster dynamics, while the evolution of the quasi-star is another important key step of the runaway merger scenario. Studies have found that quasi-stars' mass ($M_{\rm q}$ hereafter) can reach up to $10^6~M_{\odot}$ in principle (given infinite ``fuel''), and the remnant BH mass is $M_{\rm BH} \sim 0.1M_{\rm q}$ \cite[e.g.,][]{begelman_evolution_2010,ball_structure_2011}. These models also find the quasi-stars' lifetime to be $\sim 3~{\rm Myr}$, with only a weak dependence on their masses.
  
These models, taken at face value, however, would actually imply almost no IMBHs in GCs or other dense stellar systems. The problem is that most GCs (let alone nuclear stellar clusters or galaxy bulges) have half-mass relaxation time scale much longer than 100 Myr (e.g., $t_{\rm rlx}\sim$ 2.5 Gyr for M15). However, the studies cited above assumed the \emph{initial} mass profile of GCs was essentially identical to the mass profiles of nearby relaxed clusters observed today (e.g., with a flat King-type central density profile). In short, if one were to assume that the GCs' present-day mass distribution reflects their mass distribution at formation, this would rule out the runaway merger channel in most globular clusters. However, calculations following the dynamical evolution of globular clusters over cosmological timescales unanimously find that this is not a good assumption \citep{giersz_2013_mocca,wang_2016_dragon, baumgardt_2018_gcs, kremer_2019_cmc}. Rather, the combined effects of stellar evolution and mass loss, dynamical ejections, mass segregation and ``binary burning", and tidal heating/stripping all tend to puff up and flatten the central mass profile slope of dense stellar systems (usually on time scales far shorter than the $N$-body relaxation time), implying that many presently-observed clusters once had much denser inner cores.
  
Indeed, the closest observable cousins to proto-globular clusters, young massive clusters (YMCs), are generally found to have density profiles that are significantly different from old globular clusters of comparable mass. Their half-mass radii are generally smaller, with a typical half-mass radius of $\sim 1\,\rm pc$ that has no clear correlation with mass \citep{ryon:2015.m83.clusters,ryon:2017.ymc.profiles}, and thus their relaxation times are generally shorter. They also have a relatively compact density profile with an outer asymptotic power-law slope $\rho \propto r^{-\eta_1}$, where $\eta_1$ is typically in the range $2-3$ \citep{grudic_top_2018}. \citet{grudic_top_2018} further found that hydrodynamical simulations of YMC formation were able to reproduce this density profile robustly, and proposed that these density profiles arise from the star cluster assembly process.
  
In this article we revisit the basic physical processes involved in the assembly a massive stellar object in the centre of a dense star cluster, using the results of the \citet{grudic_top_2018} simulations, which successfully reproduce observed YMC outer density profiles \citep{ryon:2015.m83.clusters,ryon:2017.ymc.profiles}, as well as a range of giant molecular cloud (GMC) properties including their turbulent structure, magnetic field strengths, and stellar auto-correlation functions or stellar clustering \citep{grudic_2020_gmc_cosmic}. These simulations attempt to capture (to the extent possible with state-of-the-art simulations) the cluster properties as they form, which is the most relevant time for potential IMBH formation. Using these simulation results to guide our space of cluster models, we perform a set of Monte Carlo (MC) simulations to track the mass segregation process and study the evolution of the central mass (the quasi-star). Using these methods we predict the mass growth history of the quasi-star and its dependence on the properties of the progenitor cloud or host cluster.

The article is organized as follows: in Section \ref{sec:methods} we introduce the analytical and numerical methods used to study mass segregation and the runaway growth of massive objects in star clusters; in Section \ref{sec:result}, we show the numerical results from the MC simulations and discuss some secondary effects; in Section \ref{sec:discussion}, we expand the discussion to observational aspects and make predictions; finally, in Section \ref{sec:conclusions} we summarize our main findings.

\section{Models and methods}
\label{sec:methods}

\begin{figure*}
    \includegraphics[width=.75\textwidth]{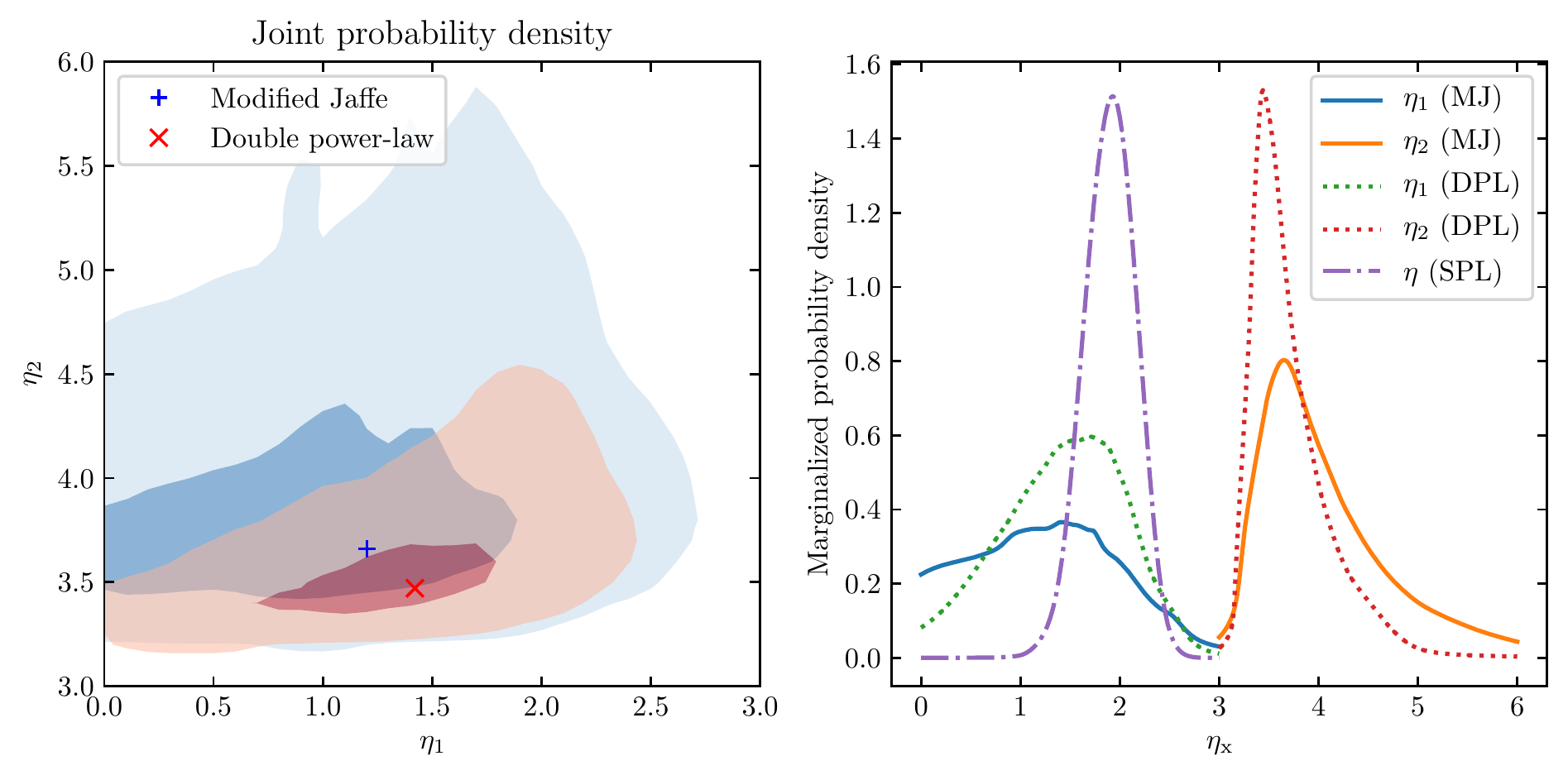}
    \vspace{-0.4cm}
    \caption{Probability distribution function (PDF) of the best-fit analytic density profile innert slopes ($\eta_{1}$, where $\rho\propto (r/r_{\rm h})^{-\eta_{1}}$ as $r\rightarrow 0$) and outer slopes ($\eta_{2}$, relevant as $r\rightarrow \infty$), fit to our entire library of MHD star cluster formation simulations. We compare both ``modified Jaffe'' (Eq.~\ref{equ:jaffe}; MJ) and ``double power-law'' (Eq.~\ref{equ:doublepower}; DPL) and (at {\em right}) single power-law (SPL) fits to just $r < r_{\rm h}$.
    \textbf{Left}: Joint PDF of $\eta_{1}$ and $\eta_{2}$. Darker (lighter) counters denote the $1~\sigma$ ($2~\sigma$) inclusion contours, while crosses show the local maxima. 
    \textbf{Right:} Marginal 1D PDFs for $\eta_{1}$ and $\eta_{2}$. 
    The MJ, DPL, and SPL fits are statistically consistent, though MJ shows larger covariance between $\eta_{1}$ and $\eta_{2}$ owing to the much less-sharp ``knee''; SPL has no covariance by construction. 
    Independent of fitting methodology, the simulations clearly exhibit steep inner profiles at formation with most-common $\eta_{1} \sim 1.5-2$, closer to isothermal ($\eta_{1}=2$) than to their post-relaxation King-like ($\eta_{1}=0$) profiles.}
    \label{fig:dist-eta}
\end{figure*}

\subsection{Initial Conditions from Cluster Formation Simulations}

For our initial conditions, we extract catalogues of star clusters as they form in the simulations from \cite{grudic_when_2018,grudic_top_2018}. These are N-body plus magneto-hydrodynamic (MHD) simulations of cloud collapse and star formation, including detailed models for radiative cooling and chemistry, star formation, and ``feedback'' once stars form in the form of radiation (e.g. radiation pressure and HII regions), stellar winds, and supernovae. The simulations follow the collapse of giant molecular clouds, the assembly of star clusters, and the eventual dispersal of gas due to stellar feedback. One such simulation of e.g. a massive complex can produce many independent clusters: we identify gravitationally-bound star clusters remaining after gas dispersal\footnote{We have also compared the results extracting clusters at the time of peak star formation; the time difference is sufficiently small that it has little effect on our results.} using a group-finder which associates stars belonging to a common potential well which are also gravitationally bound within that well (see Appendix A in \citet{grudic_top_2018} for details). We restrict to clusters which form $>100$ bound star particles.

This gives us an ensemble of $\sim1000$ clusters ``at formation,'' one of which is shown in Fig.~\ref{fig:fit-evolution} Note that this sample of clusters should not be considered statistically representative of a cluster population that would form in a real galaxy: the initial conditions of the simulations were simply uniformly sampled on a logarithmic grid in mass-radius parameter space, which is ideal for our study here. 

These simulations are designed to (1) sample an enormous parameter space, and (2) simulate large complexes through the entirety of star formation and stellar evolution: as such, the numerical resolution is such that individual ``star particles'' in the original simulations represent an IMF-averaged ensemble of stars. To properly evolve stellar dynamics, we therefore re-sample each star particle into an ensemble of individual stars, drawing probabilistically from the stellar initial mass function (IMF) conserving total stellar mass. By default (since it is the same used for the original simulation stellar evolution models) we adopt a \cite[]{kroupa_variation_2001} IMF with an upper mass limit of $m_{\rm max}=100~M_{\odot}$.\footnote{We have tested and adopting instead a \citet{chabrier:imf} makes negligible difference to our conclusions. Likewise we find that varying the upper ``cutoff'' mass of the IMF makes only weak (logarithmic) corrections to our predictions (because these stars contribute negligibly to the total massive-star stellar mass budget).} For this assumption the median stellar mass is $\langle m \rangle \approx 0.38~M_{\odot}$, and the mean is $1.5~M_{\odot}$.

\subsection{Analytic Models from the Simulations}

Although our simulation suite is extensive, it is still limited by (1) finite sampling of parameter space and (2) finite resolution. Especially in cluster centers (particularly important here), the original simulation will always produce finite-resolution effects. Moreover, although the simulated clusters have some non-axisymmetric structure, we generally find this is small and generates torques which are weak compared to dynamical friction (discussed below). Therefore, it is especially useful to also consider general analytic models for the initial conditions, {\em motivated} by the cluster catalogue from our simulation suite. 

We consider three simple, spherically-symmetric analytic density profiles, which we will show allows us to capture almost all of the key behaviors we study. These are shown in Fig.~\ref{fig:fit-evolution} as fits to one example profile.\footnote{We have experimented with a variety of different methods for fitting the analytic profiles to the simulation outputs, and find the most robust results fitting directly to the spherically-averaged $\rho(r)$ in log-log space with uniform weights but constraining the analytic fit to reproduce the total mass and half-mass radius (specifying $\rho_{c}$ and $r_{\rm c}$) exactly, so only the slopes $\eta_{1,2}$ are ``free.''} First, a ``Modified Jaffe''  model (from \citealt{binney_galactic_2011}, Eq.~2.64): 
\begin{align}
    \rho(r) = \rho_{\rm c}\left(\frac{r}{r_{\rm c}}\right)^{-\eta_1}\left(1+\frac{r}{r_{\rm c}}\right)^{-\eta_2+\eta_1}.\label{equ:jaffe}
\end{align}
with inner power-law slope $\eta_1$, outer slope $\eta_2$, turnover radius $r_{\rm c}$, and normalization $\rho_{c}$ (given by e.g.\ the total mass). We also consider a similar ``double power'' law model: 
\begin{align}
    \rho(r) =& \frac{\rho_{\rm c}}{(r/r_{\rm c})^{\eta_1}+(r/r_{\rm c})^{\eta_2}}, \label{equ:doublepower}
\end{align}
which has the same qualitative features as the ``Modified Jaffe'' model but features a much sharper turnover around $r_{\rm c}$, which is useful in what follows as it dramatically reduces the covariance between the parameters $\eta_1$ and $\eta_2$. 

Finally, we also consider a ``single power'' law model: $\rho(r)=\rho_{\rm c}(r/r_{\rm c})^{-\eta}$ for $r<r_{\rm c}$. This obviously cannot fit any mass profile over the entire dynamic range of $r$ with finite mass; we therefore restrict the fit only to radii smaller than the half-mass radius (so $\eta \approx \eta_{1}$). This is included here because it allows us to derive some simple analytic expressions in regimes where the inner profile dominates the behavior.

\begin{figure*}
    \includegraphics[width=.8\textwidth]{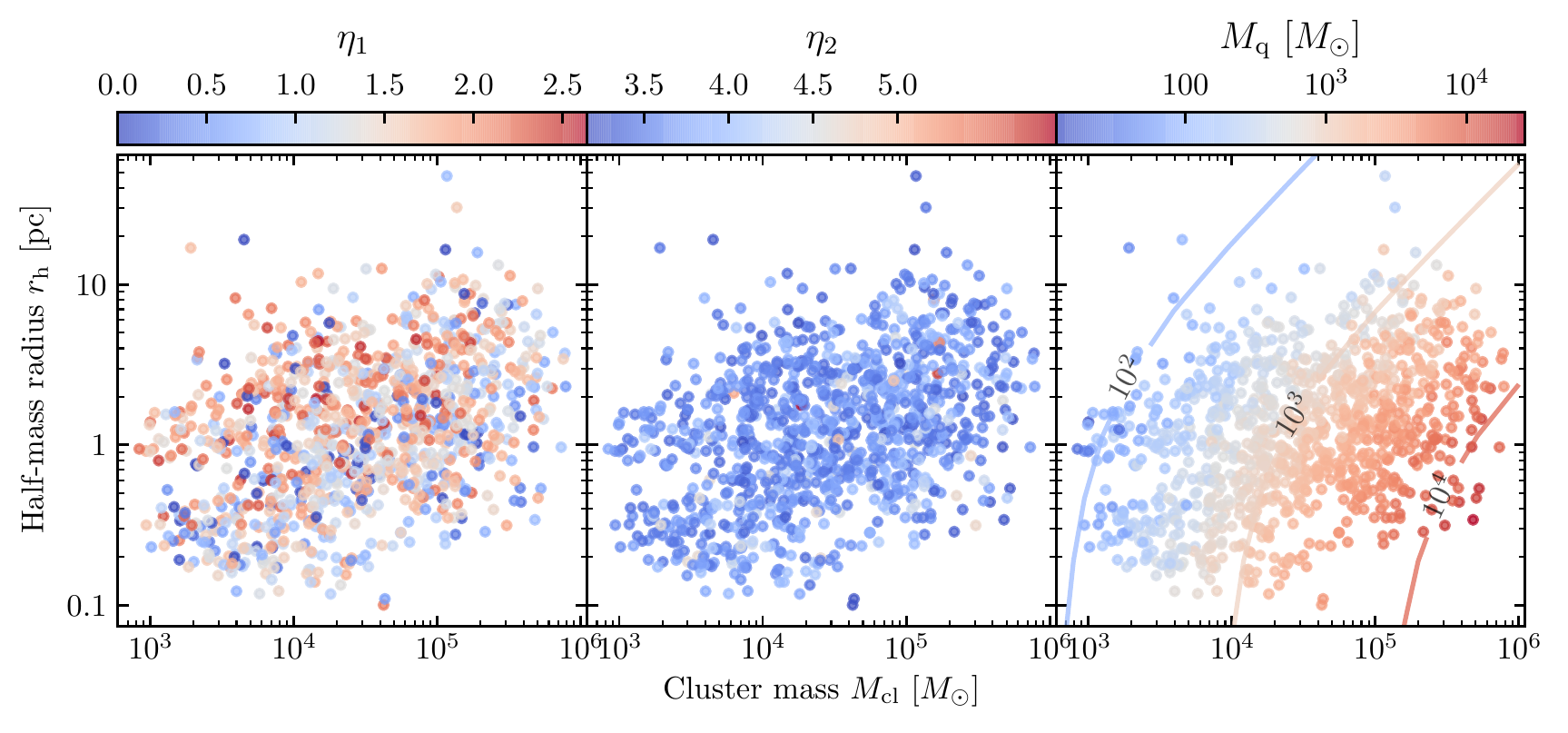} 
    \vspace{-0.3cm}
    \caption{Properties and predictions from our cluster catalog with respect to cluster mass $M_{\rm cl}$ and effective radius ($r_{\rm h}$). 
    \textbf{Left}: Best-fit inner density profile slope parameter ($\eta_1$).
    \textbf{Middle}: Best-fit outer density profile slope parameter ($\eta_2$).
    Notably, there is no significant systematic dependence of the density profile {\em shape} parameters $\eta_{1}$ or $\eta_{2}$ on the cluster mass and radius, consistent with scale-free predictions for substructure in turbulent gravitational fragmentation \citep{guszejnov:universal.scalings}. 
    \textbf{Right}: Predicted central object mass $M_{\rm q}$, measured as $M_{\rm c}(t=t_{\rm c}$, i.e.\ at the ``crossover time'' as in e.g. Fig.~\ref{fig:fit-evolution}, from our ``full model,'' as a function of $M_{\rm c}$ and $r_{\rm h}$. We compare the predicted values (contours, with $M_{\rm q}/M_{\odot}$ labeled) if we assume all clusters have an identical universal double power-law mass profile with the ``typical'' values of $\eta_{1}=\eta_{1*}$ and $\eta_{2}=\eta_{2*}$ (\S~\ref{sec:density.profiles}). This reproduces the results well, indicating that sub-structure in the simulations, and variation in mass profile shape from cluster-to-cluster, do not strongly influence our conclusions.
    \label{fig:m-r-scatters}}
\end{figure*}

\subsection{Sinking and ``Merging'' Stars}
\label{sec:method:sink.merge}

Even with the simplifications above, integrating the full N-body dynamics of massive stars through a cluster into merger with a central object is computationally impossible (both given our large parameter space of models and sample of ``clusters'' reaching $\sim 10^{8}\,M_{\odot}$, let alone uncertainties in the actual size/evolution of the central object). However, full N-body studies of a small number of smaller clusters \citep[e.g.][]{portegies_zwart_runaway_2002,gurkan_formation_2004,alessandrini_2014_df_cluster} have shown that dynamical friction is an excellent approximation to rate of sinking and merger of massive stars with $m\gg \langle m\rangle$ (which are those that dominate the buildup of a central quasi-star on the timescales of interest). This quickly circularizes the orbits of the massive stars and leads to orbital decay with 
\begin{align}
  \dot{r} =& -4\pi\,G^{2}\,m\,\ln{\Lambda}\,v_{c}^{-3}\,\rho_{b}\,r \label{eqn:rdot}
\end{align}
where $m$ is the mass of the sinking star, $v_{c}^{2} \equiv G\,M(<r)/r$ reflects the enclosed mass $M(<r)$ inside $r$, $\rho_{b}$ is the density of the background stars at radius $r$ (e.g.\ the $\rho(r)$ in the profiles above), and $\Lambda$ is a Coulomb logarithm which we take to be $\Lambda(r) \approx 0.1\,M(<r)/\langle m \rangle$. Less massive stars will not sink: we approximate this (conservatively, for now) by simply applying a cutoff ignoring any $\dot{r}$ below $m_{\rm min}=8\,M_{\odot}$.\footnote{This ignores back-reaction causing lower-mass stars to migrate outwards, but this is a small effect on the timescales we consider, and we show below the exact choice of $m_{\rm min}$ also has relatively weak effects on our conclusions.} 

From a MC realization of the ICs (either directly from the formation simulations, or analytic fits), we then evolve the system forward in time. As massive stars approach the center, the first to reach the center (region interior to which there are no other $m>m_{\rm min}$ stars) becomes the ``seed'' quasi-star (with mass $M_{c}$). Because we are interested in mergers {\em while the massive stars are on the main sequence}, we subsequently ``merge'' into this any massive star which (1) has not yet reached the end of its main sequence lifetime (adopting the relation from \citealt{mottram_rms_2011}; typically $\sim3$\,Myr for the most massive stars), (2) approaches the quasi-star within a radius $r < r_{q}$ which represents some ``interacting binary'' or ``common envelope'' radius (for which we take the value quoted by \citet{hosokawa_formation_2013} for models of a rapidly-accreting protostar: $r_{q} \approx 2600\,R_{\odot}\,(M_{q}/100\,M_{\odot})^{1/2}$),\footnote{This is essentially an extrapolation from ``normal'' pre-main sequence stars. Of course the sizes of quasi-stars are purely theoretical and uncertain: however varying this by factors of several has very little effect on our conclusions, as the ``sinking'' times around these radii are relatively small. But we need to include some finite ``merger radius'' since we do not model effects like gravitational wave emission which could merge point-mass-like particles.} and (3) reaches before the quasi-star itself has reached the end of its lifetime. We simply add the merged mass to $M_{c}$, neglecting e.g. mass-loss associated with the merger.
Note that during collisions the central object mass $M_c(t)$ will contribute to the total enclosed mass $M(<r)$ as an additional point mass, which is included as a correction when solving Eq. \eqref{eqn:rdot} in our \emph{full} version of MC simulations. This ``point mass correction'' (PMC) is not included in our analytical calculation and the corresponding MC simulations.

The quasi-star ``lifetime'' essentially sets the end of our simulation, and the final mass of the quasi-star. Studies of quasi-star structure \cite[e.g.,][]{goodman_supermassive_2004,schleicher_massive_2013,ball_structure_2011} have found that because these stars are approximately Eddington-limited, they have lifetimes $\sim3$\,Myr akin to massive stars. We have therefore considered simply taking the mass $M_{q} = M_{c}(t=3\,{\rm Myr})$. We have also considered a more sophisticated model motivated by the same pre-main-sequence models described above: some accretion rate $\dot{M}_{\rm acc} = d M_{c}/d t$ from mergers sustains the quasi-star lifetime and keeps it ``puffed up'' (allowing efficient mergers) as long as it is larger than the fuel consumption/loss rate from a combination of nuclear burning and stellar mass-loss, which occurs on a characteristic timescale $t_{0}\sim 3\,{\rm Myr}$. We therefore take the final $M_{q}$ to be $M_{c}$ at the first time where $\dot{M}_{\rm acc}$ falls below $M_{c}/t_{0}$. In practice, because the merging stars also have lifetimes $\sim 3\,$Myr, it makes very little difference which of these assumptions we adopt.

\begin{figure}
    \includegraphics[width= \linewidth]{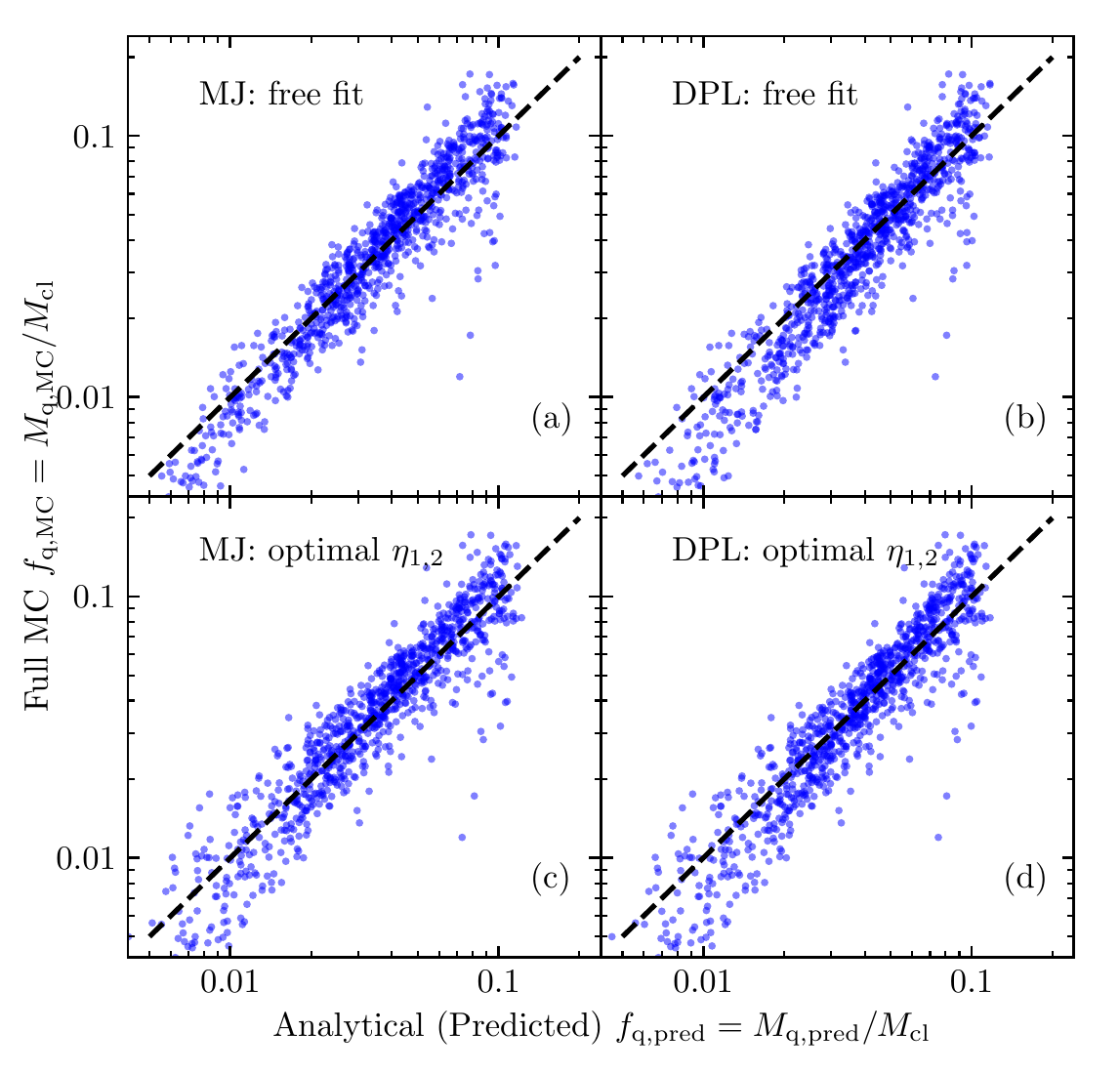}
    \vspace{-0.8cm}
    \caption{Quantitative comparison of the maximum central object mass $M_{q}$ predicted by different analytic models for the density profile, to that calculated using the full simulation 3D density information, for all $\sim1000$ MHD star-formation simulation clusters in our library. We normalize both by cluster mass $M_{\rm cl}$ to reduce the dynamic range and better highlight any discrepancies. Dashed lines show identity (MC equals analytic). 
    \textbf{(a)}: Results assuming an analytic modified Jaffe profile with all clusters fitted separately with 4 free parameters.
    \textbf{(b)}: Double power model with all clusters fitted separately with 4 free parameters.
    \textbf{(c)}: Modified Jaffe model assuming a universal profile shape with the slopes $\eta_{1*}$ and $\eta_{2*}$ (but allowing $M_{\rm cl}$ and $r_{\rm h}$ to vary, matched to the exact simulation values for each simulation).
    \textbf{(d)}: Double power model assuming a universal profile shape with $\eta_{1*}$ and $\eta_{2*}$.
    In general, assuming universal, smooth, analytic, 1D density profiles introduces relatively small errors into our estimates of $M_{\rm q}$ ({\em provided} we adopt the correct ``at formation'' slopes), suggesting it is reasonable to apply these to YMCs and other young objects for which $M_{\rm cl}$ and $r_{\rm h}$ can be measured but $\rho(r)$ as $r\rightarrow 0$ cannot be resolved.
    \label{fig:th-exp}}
\end{figure}

\section{Results}
\label{sec:result}

\subsection{Density Profiles}
\label{sec:density.profiles}

\citet{grudic_top_2018} showed that the simulated clusters here produce a distribution of density profile shapes {\em after relaxation} in good agreement with observations; however, no analysis of the {\em inner} density profiles {\em at formation} was performed. In Fig.~\ref{fig:dist-eta}, we show the distribution of the inner ($\eta_1$) and outer ($\eta_2$) mass profile slopes fit to all clusters. In both ``modified Jaffe'' and ``double-power-law'' models, the outer slopes are typically in the range of 3.5-4.5 as found in \citet{grudic_top_2018}. The inner slopes cluster around $1-2.5$ (as compared to post-relaxation profiles, which broadly follow a ``flat'' \citealt{EFF1987} distribution). The best fit distribution for $\eta_1$ is most narrowly-peaked (around $\eta_1\approx 2$) for the ``single-power-law'' fits (fit to just $r$ within the half-mass radius), and most broad for the ``modified Jaffe'' fit. Our extensive experimentation with different fitting methods indicates that this directly traces the covariance between $\eta_1$ and $\eta_2$. The single-power fit, with only one slope, has no $\eta_1-\eta_2$ covariance. The double-power fit, with a ``sharp'' break, has weak covariance between $\eta_1$ and $\eta_2$ which ``smears'' the best-fit $\eta_1$. The modified Jaffe fit exhibits very strong covariance between $\eta_1$ and $\eta_2$, with a wide range of allowed fits for any given simulation profile.\footnote{It is important to note that because of the covariance in the fits, models with the modified Jaffe fit with $\eta_1\sim 0$ give $r_{c} \ll r_{\rm half}$, i.e.\ the ``rollover'' occurs very slowly down to extremely small radii (often below the simulation resolution) -- so the central densities are still large.}

\begin{figure*}
    \includegraphics[width=.75\linewidth]{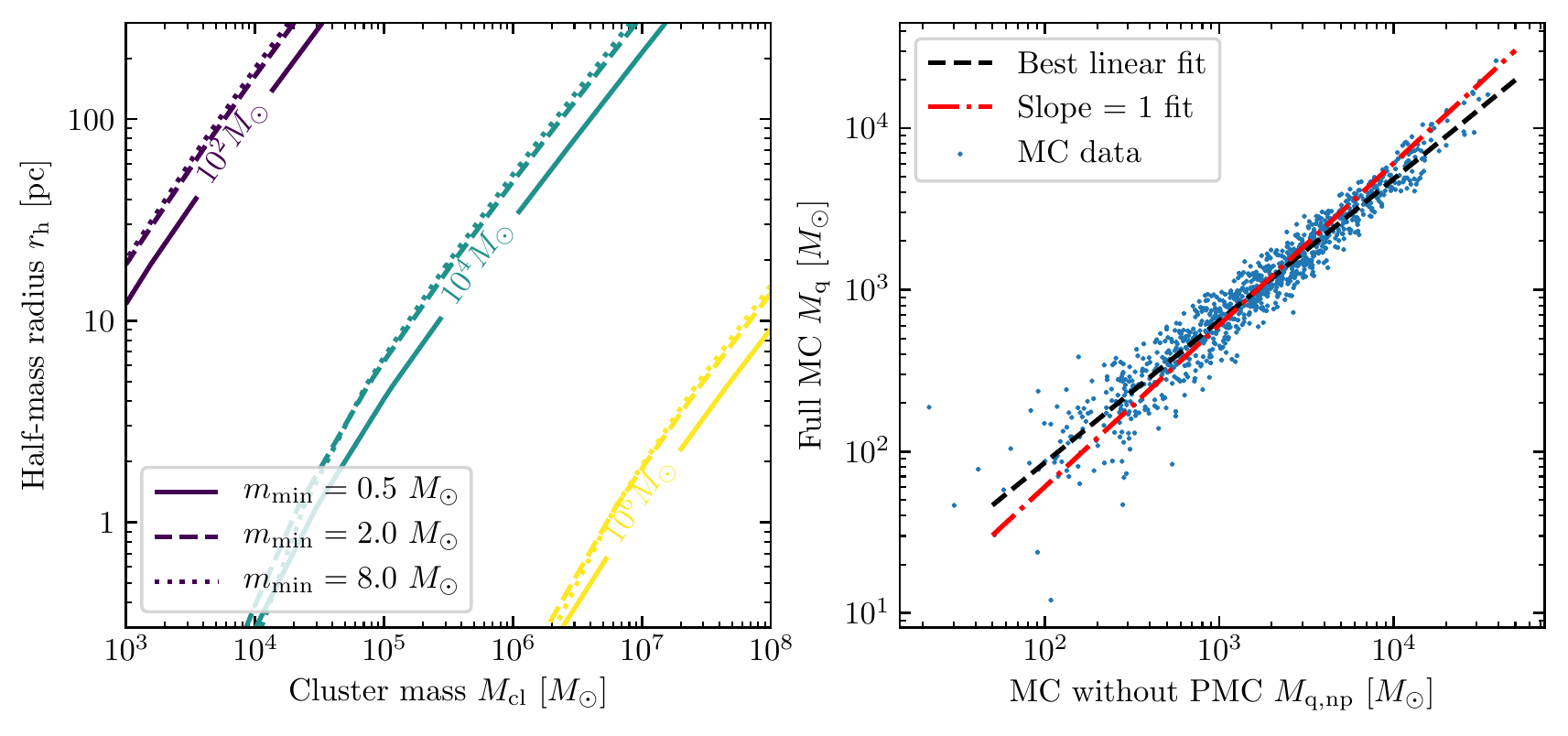}
    \vspace{-0.3cm}
    \caption{The effect of the ``lower mass cutoff'' (mass limit where dynamical friction remains a good approximation) and point-mass corrections on quasi-star masses $M_{\rm q}$. 
    \textbf{Left}: Varying $m_{\rm min}$, the minimum stellar mass where we assume that a dynamical-friction-type orbital decay (Eq.~\ref{eqn:rdot}, requiring stellar masses $m \gg \langle m \rangle \sim 0.38\,M_{\odot}$) is valid. Lines show the predicted $M_{\rm q}$ as a function of cluster mass $M_{\rm cl}$ and size $r_{\rm h}$, assuming the universal best-fit $\eta_{1*}$, $\eta_{2*}$, for each $m_{\rm min}$, and otherwise adopting our ``full'' model. This produces nearly-negligible differences, as stars with masses $\ll 8\,M_{\odot}$ sink inefficiently even if dynamical friction were a good approximation for their dynamics. The effects of varying the upper mass-limit of the IMF from $100\,M_{\odot}$ (not shown) are also negligible.
    \textbf{Right}: Effect of ignoring the ``point mass correction'' (not accounting for the finite N-body effect of the quasi-star itself, see Fig.~\ref{fig:fit-evolution}). Even if we ignore these corrections, we obtain $M_{\rm q,\,np}$ very similar to our full model $M_{\rm q}$, just systematically larger by a modest factor -- a linear fit gives $M_{\rm q} \approx 0.6\,M_{\rm q,\,np}$ (fitting an arbitrary power-law gives $(M_{\rm q}/2000\,M_{\odot}) \approx 0.6\,(M_{\rm q,\,np}/2000\,M_{\odot})^{0.9}$, but the difference from the ``slope $=1$ fit'' is not significant). 
    Accounting for these finite N-body effects produces a not-negligible correction to $M_{\rm q}$, but it is largely a systematic effect which does not change our qualitative conclusions.
    \label{fig:mcut}}
\end{figure*}
Fig.~\ref{fig:m-r-scatters} shows that the best-fit $\eta_{1}$ and $\eta_{2}$ do {\em not} depend systematically on either cluster half-mass radius $r_{\rm h}$ or mass $M_{\rm cl}$. Likewise our cluster catalogue includes simulations with progenitor clouds of different metallicities ($Z/Z_{\odot}=0.01-1$), and we see no dependence on $Z$. 

Even though the exact values of the ``inner slope'' $\eta_1$ can vary between fits, the fact that these are relatively narrowly constrained (within covariances, i.e.\ all are ``good'' representations of the data), and that they do not depend systematically on cloud mass/size, means that we obtain reasonably good estimates for the final central object mass from our full MC calculation, using an idealized mass profile fit with either of the fitting functions {\em or} assuming a ``universal'' mass profile shape across all clusters (Fig.~\ref{fig:th-exp}). 
The modal fit values for ($\eta_1$, $\eta_2$) are ($\eta_{\rm 1M}$, $\eta_{\rm 2M}$) = (1.2, 3.7) for the modified Jaffe, (1.4, 3.5) for the double-power-law, and (1.9) for the single power-law model. Because of the covariances, however, these are not the same as the values which give the best estimate of $M_{q}$ compared to our full MC calculation. Instead, we should ask which values ($\eta_{1\ast}$, $\eta_{2\ast}$), applied to the entire ensemble of clouds (as a ``universal'' profile shape), most accurately predict $M_{q}$ from the MC:\footnote{Formally we find the ($\eta_{1\ast}$, $\eta_{2\ast}$) which minimize the variance $\sum |\log\{M_{q,\,{\rm pred}}(\eta_{1\ast},\,\eta_{2\ast},\,r_{\rm h},\,M_{\rm cl})\} - \log\{M_{q}({\rm MC})\}|^{2}$.} these are ($\eta_{1\ast}$, $\eta_{2\ast}$)=(1.68, 4.95) for modified Jaffe,\footnote{The dramatic change in the values for modified Jaffe again indicates the covariance (with the $\eta_2=4.95$ value indicating that the outer slope plays a very small role in determining $M_{q}$).} (1.81,\,3.79) for double power-law, and (1.93) for single power-law, all of which feature a similar, isothermal-like inner slope $\eta_1 \sim 2$.

\subsection{Central Object Growth}
\label{sec:central.object.growth}

Fig.~\ref{fig:fit-evolution} shows one example of our full MC simulation, with the ensuing growth of the central object as a function of time $M_{c}(t)$. The results from the spherical analytic model with the modified Jaffe or double-power-law profiles agree very will with the full MC (the single-power-law works well up to $\sim 1-3\,$Myr, as well, where most of the accretion is from radii $\ll r_{\rm h}$). The ``crossover point'' where $\dot{M}_{\rm acc} = M_{c}/t_{0}$ occurs at $t \sim 1.3\,$Myr, but the growth rate of $M_{c}$ is slowing down already at this point, so $M_{q}$ only differs by a factor of $\sim 1.5$ if we take $M_{q}$ to be $M_{c}$ at $t=3\,$Myr, or a factor of $1.9$ if we take $M_{q}$ as $t\rightarrow \infty$. In any case, this particular cluster, chosen to be relatively extreme (with a total mass $\sim 10^{6}\,{\rm M_{\odot}}$ and initial central density of $\sim 10^{11}\,{\rm M_{\odot}\,pc^{-3}}$ at $r\lesssim 0.001\,{\rm pc}$) is able to merge most of its massive stars ($\sim 10\%$ of the total stellar mass) within $<3\,$Myr. 

Fig.~\ref{fig:m-r-scatters} shows $M_{q}$ from our full MC calculation for each simulated cluster, as a function of cluster mass and half-mass radius. There is a clear trend where more massive clusters $M_{\rm cl}$ at the same size give larger $M_{\rm q}$, and a weaker but still evident trend of larger $M_{\rm q}$ for more compact clusters at fixed mass. These are expected if cluster profiles are approximately self-similar: to show this we compare the predicted $M_{\rm q}$ from analytic models with different $M_{\rm cl}$ and $r_{\rm h}$, assuming a universal density profile shape (the double power-law fit with fixed $\eta_1 = \eta_{\rm 1\ast}$,  $\eta_2 = \eta_{\rm 2\ast}$).

Fig.~\ref{fig:th-exp} compares the mass fraction which can merge to the center $f_{q} \equiv M_{q}/M_{\rm cl}$ from our full MC calculation to that obtained from the simple spherical analytic models. We compare the results from the modified Jaffe and double-power-law fits, fit individually to each simulation cluster, which predict $f_{q}$ to within $<10\%$ on average -- this indicates that deviations from symmetry, ``lumpiness'' or irregular structure in the potential and density profile, or resolution effects (e.g.\ numerical flattening or shot noise in the central density profile, as compared to the profile generated by a smooth power-law down to $r\rightarrow 0$) do not strongly influence our results. We also show the results assuming a universal profile with ($\eta_{1}$, $\eta_{2}$) = ($\eta_{1\ast}$, $\eta_{2\ast}$). This increases the scatter (as expected) but only by a modest amount: we can predict $M_{q}$ to within an rms $<0.15$\,dex assuming this universal shape at formation.

It is unclear exactly at which mass scale dynamical friction ceases to be a good approximation for the ``sinking'' of massive stars: Fig.~\ref{fig:mcut} varies the minimum mass $m_{\rm min}$ we allow to sink, to show this has only a small effect on our predictions. Varying $m_{\rm min}$ from $0.5-8\,{\rm M_{\odot}}$ changes $M_{q}$ by a factor $\sim 2$, because (a) lower-mass stars sink more slowly (even if we allow them to sink), and (b) the Salpeter IMF is not extremely steep, so the total mass of stars ``sinking'' only changes with $m_{\rm min}^{-0.3}$. 

Because most of the mass in the IMF is not in the highest-mass stars, it also makes little difference if we vary the high-mass cutoff (e.g.\ changing the upper-mass cutoff of the IMF from $100\,M_{\odot}$ to $200\,M_{\odot}$ only produces a $\lesssim10\%$ difference in $M_{\rm q}$). 

Another significant uncertainty in our models is how the actual mergers/coalescence occur in the center: we simply populate stars and merge anything within some large radius in the center (reflecting the envelope size of the quasi-star), implicitly meaning there is some ``overlap'' between the envelope of the quasi-stars and our populated stars in the models. Properly determining if or how mergers once massive stars sink close to the central quasi-star requires dynamical stellar merger simulations. But even within our simple model, stars can still ``stall'' near the center. In Fig.~\ref{fig:mcut}, we consider a model variation where we simply merge any star which reaches the radius where the proto-star {\em would} dominate (be more than $1/2$ of) the enclosed mass $M(r)$, and ignore the mass of $M_{q}$ itself in calculating $v_{c}$ in Eq.~\ref{eqn:rdot}: these changes essentially guarantee that any massive star which approaches small $r$ merges. We see that this systematically increases $M_{q}/M_{\rm cl}$, as expected, by a factor $\sim 2$. This in turn means that our ``default'' model is predicting an order-unity fraction of massive stars near $r\sim 0$ ``stall'' or otherwise fail to merge, a reasonable order-of-magnitude approximation to few-body simulations. It also implies that these correction does not change our qualitative conclusions.

\begin{figure*}
    \includegraphics[width=.85\textwidth]{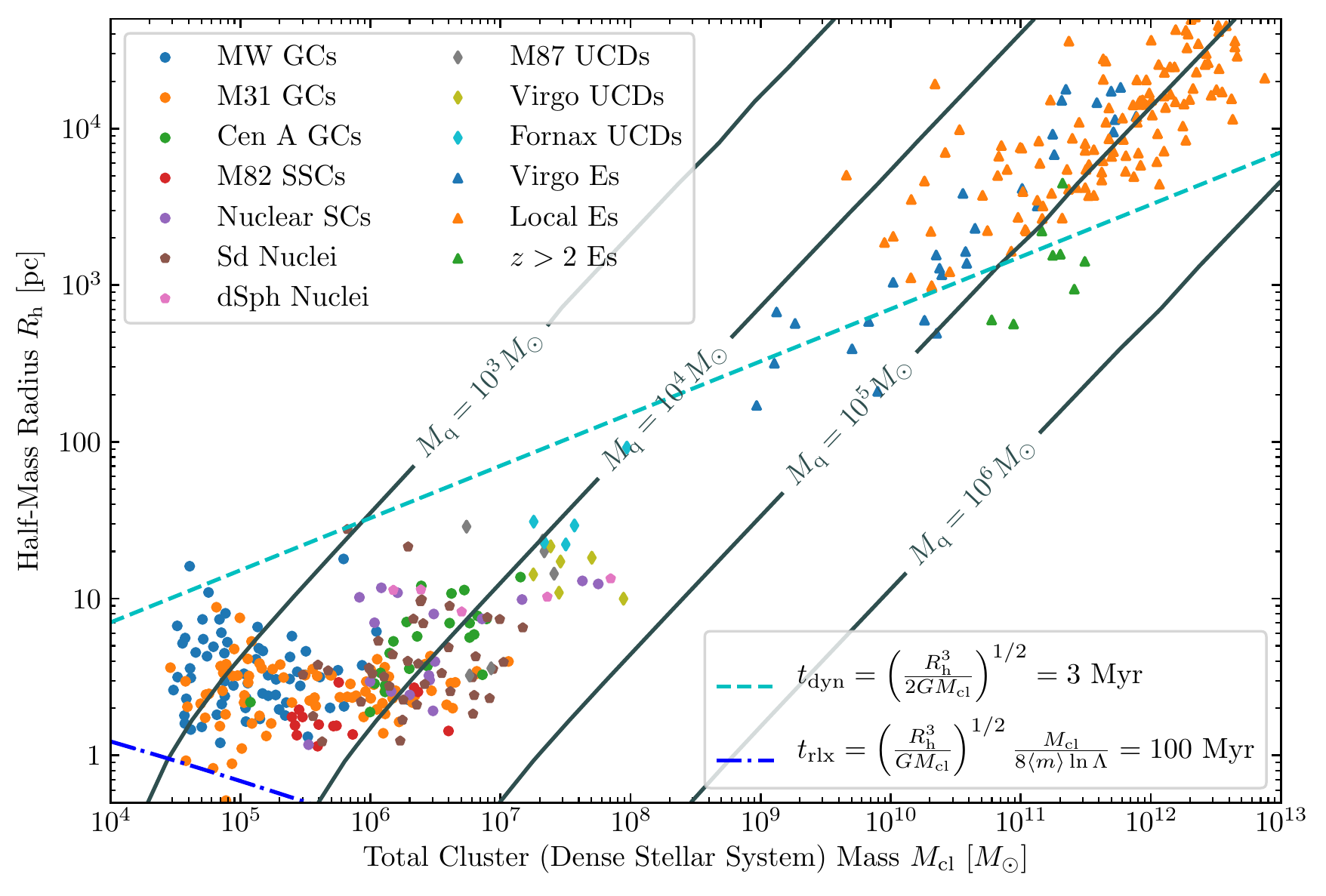}
    \vspace{-0.3cm}
    \caption{Predictions for the peak central object mass $M_{q}$ given by our full models assuming the best-fit universal at-formation density profile parameters $\eta_{1*}$, $\eta_{2*}$, and given cluster mass $M_{\rm cl}$ and projected 2D circular half-mass radius $R_{\rm h}$ (solid lines label contours of constant $M_{q}$). We note where the dynamical time $t_{\rm dyn} = 3\,$Myr (systems with $R_{\rm h}$ above this line have $t_{\rm dyn} \gg 3\,$Myr, making our assumption that the stars are approximately co-eval on the timescales of interest for quasi-star evolution suspect). We also compare the contour below which the cluster-scale N-body relaxation time $r_{\rm rlx} < 100\,$Myr where ``complete core collapse'' (requiring full N-body simulations) as opposed to sinking in the center should occur (almost no clusters meet this criteria). We compare the masses and radii of a variety of {\em observed} dense stellar systems (references in text; \S~\ref{sec:discussion}), including globular clusters (GCs), super star clusters (SSCs), nuclear star clusters (NSCs) and dwarf galaxy stellar nuclei, ultra-compact dwarf galaxies (UCDs) and the elliptical galaxies (Es). If we assume their masses and radii have not evolved dramatically since formation, this gives a rough prediction for their initial quasi-star masses. Although $M_{\rm q}$ could reach as large as $\gtrsim 10^{4}\,M_{\odot}$ in the most massive SSCs/nuclei/UCDs, this is generally much smaller than the present-day SMBH masses detected in such systems (even moreso for the $M_{\rm q} \gtrsim 10^{5}\,M_{\odot}$ in the most massive Es, but these also have $t_{\rm dyn} \gtrsim 3\,$Myr). 
    \label{fig:sweetspot}}
\end{figure*}

\section{Comparison to Previous Work}

As discussed in \S~\ref{sec:intro}, \cite{portegies_zwart_runaway_2002} and \cite{gurkan_formation_2004} considered detailed $N$-body simulations to follow runaway merging, but used adopted very different mass profile shapes (similar to post-relaxation clusters today). {\em If} we adopt a similar profile shape to their default (e.g.\ a Plummer-like ($\eta_1$, $\eta_2$) = (0 , 5)), and then run our full model to calculate $M_{q}$ for a wide range of $M_{\rm cl}$ and $r_{\rm h}$ (sampling the values of our simulation library) then -- despite the other  simplifications here -- we obtain quite good agreement (within a factor of a few) with {\em both} their requirement that the cluster must have $t_{\rm rlx} \lesssim 25\,$Myr to produce any appreciable growth of $M_{\rm c}$ and the peak mass $M_{\rm q}$ or resulting mass fraction $f_{q}=M_{q}/M_{\rm cl}$ in a central massive object produced when this criterion is met. This is reassuring, and implies our methodology is reasonable. The key difference in our predictions, compared to theirs, arises because our MHD star formation simulations predict quite different values of $\eta_{1}$ compared to those they considered.

Some other recent studies have considered runway stellar mergers in initial conditions closer to those here (but with a more limited or ad-hoc choice of initial profiles). \citet{sakurai_formation_2017} adopted a similar approach to that here, using hydrodynamic simulations to select dense ($M_{\rm cl} \sim 10^{5}\,M_{\odot}$, King-profile core $r_{\rm c}\sim 0.4\,$pc) proto-galactic ``clouds'' and then using those to set up initial conditions for N-body simulations: although our survey is intended to match much later-forming star clusters, where the cloud properties overlap we find similar $M_{q}$ within a factor $\sim 2$ for each of the $\sim 8$ clusters they simulate (assuming a typical $r_{\rm c}/r_{\rm h} \sim 0.1$). 

In parallel, \citet{devecchi_formation_2009} considered analytic models for cloud/cluster formation, to estimate typical cloud densities in the early universe, coupled to a simple prescription from \citet{portegies_zwart_runaway_2002} for the fraction of clusters undergoing runaway, to argue $f_{q}$ could reach $\sim 0.05$ for dense clusters formed in the early Universe because these produce steep central profiles ($\eta_{1} \gtrsim 5/3$), broadly similar to our conclusions. And recently, \citet{tagawa_making_2019} performed semi-analytic calculations qualitatively akin to those here, considering a much more limited range of profiles (but taking steep $\eta_{1}$) but much more detailed models for the (proto)-stellar evolution of the quasi-star and merger criteria, but conclude that effective growth ceases at $\sim 3\,$Myr (as we assume here) with a similar effective radius for merger (versus $M_{q}$) as we adopt here. More recently, \citet{rizzuto_intermediate_2020} performed a series of $N$-body simulations of YMCs based on King-type profiles, indicating that massive stars weighing up to $\sim 400~M_\odot$ may form within $5\textrm{--}15~{\rm Myr}$ and sequentially become IMBHs. The results generally support our semi-analytical model, which is based on a more realistic parameter space of YMC density profiles and limited lifetime for quasi-stars.

\section{Discussion}
\label{sec:discussion}

We now consider the implications of our results for real dense stellar systems. Fig.~\ref{fig:sweetspot} plots the distribution in $M_{\rm cl}$ and $r_{\rm h}$ of a wide variety of known dense, stellar-dominated, dispersion-supported systems: globulars and YMCs, super-star clusters (SSCs), nuclear star clusters in different dwarf and late-type galaxies (NSCs), ultra-compact dwarf galaxies (UCDs), nearby and high-redshift compact elliptical galaxies and bulge-dominated galaxies (Es). The sizes and masses are compiled in \citet{hopkins:maximum.surface.densities}, from observations by \citet{harris96:mw.gcs,barmby:m31.gcs,rejkuba:cenA.gcs,mccrady:m82.sscs,walcher05:nuclei.mdyn,boker04:nuclei.scalings,geha02:dE.nuclei,hasegan:M87.ucds,evstigneeva:virgo.ucds,hilker:fornax.ucds,jk:profiles,lauer:bimodal.profiles,vandokkum:z2.sizes}. We have no way of knowing their properties ``at formation,'' but because our full simulations can be reasonably approximated by assuming a ``universal'' profile shape at formation, we compare the contours of $M_{q}$ predicted by our model with a universal at-formation ($\eta_{1\ast}$, $\eta_{2\ast}$). This does assume that the total mass and size have not evolved much since formation, an obviously uncertain assumption, but likely plausible since most of these systems have N-body relaxation times longer than the Hubble time.

From the detailed studies of quasi-star evolution noted in \S~\ref{sec:intro}, we will also assume in what follows that any quasi-star leaves behind a ``relic'' BH of mass $M_{\rm BH} \sim 0.1\,\epsilon_{0.1}\,M_{q}$. This ``fudge factor'' $\epsilon_{0.1}$ accounts for processes including inefficiency of final mergers, mass loss/ejection during merges, stellar winds and mass loss during collapse from the quasi-star.

\begin{table*}
    \centering
    \begin{tabular}{lrrcl}
        \toprule
           Object & ${M_{\rm cl}}/{M_{\odot}}$ & ${r_{\rm h}}/{\rm pc}$ & $\frac{M_{\rm BH}}{\epsilon_{0.1}\,M_{\odot}}$ (predicted) & $M_{\rm BH}/M_{\odot}$ (observed) \\
        \midrule
            $\omega$ Cen [1] & $3.2\times10^6$ & 8.4 & 600 & 40,000 [1,8]; <12,000 [9]\\
            47 Tucanae [3]  & $1.1\times10^6$ & 8.2 & 300 & 2,300 [13]; <1,700 [3,7] \\
            G1 (M31) [4]  & $7.6\times10^6$ & 6.8 & 1500 & 17,000 [14]; no evidence ($\lesssim 20,000$) [4]\\
            M3 [15]  & $2.7\times10^5$ & 3.4 & 200 &  <5,300 [15]\\
            M13 [15]  & $3.0\times10^5$ & 9.5 & 100 &  <8,600 [15]\\
            M15 [6]  & $6.5\times10^5$ & 7.7 & 200 &  <500 [16]\\
            M92 [15]  & $2.3\times10^5$ & 2.6 & 200 &  <1,000 [15]\\
            NGC 1851 [18]  & $3.7\times10^5$ & 2.4 & 300 &  <2,000 [18]\\
            NGC 1904 [18]  & $1.4\times10^5$ & 4.0 & 100 &  3,000 [18]\\
            NGC 5694 [18]  & $2.6\times10^5$ & 4.4 & 100 &  <8,000 [18]\\
            NGC 5824 [18]  & $4.5\times10^5$ & 4.5 & 200 &  <6,000 [18]\\
            NGC 6093 [18]  & $3.4\times10^5$ & 3.2 & 200 &  <800 [18]\\
            NGC 6266 [18]  & $9.3\times10^5$ & 3.0 & 500 &   2,000 [18]\\
            NGC 6388 [2] & $6.8\times10^5$ & 1.5 & 700 & 28,000 [2]; <2000 [10]; 1,500 [11];  <1,200 [12] \\
            NGC 6397 [6]  & $9.1\times10^4$ & 4.6 & 60 & 600 [15] \\
            NGC 6624 [5]  & $1.1\times10^5$ & 2.4 & 100 & 7,500 [17]; no evidence ($\lesssim 10,000$) [5] \\
        \bottomrule
    \end{tabular}
    \caption{Predictions for the relic IMBH mass from runaway merging based on our study here, compared to observational estimates or upper limits for IMBH masses in well-studied clusters. Columns give: (1) Object: cluster name (with reference for its properties); (2) Cluster mass $M_{\rm cl}$; (3) Cluster half-mass radius $r_{\rm h}$; (4) Predicted relic IMBH mass $M_{\rm BH}$ from our models (see e.g. Eq.~\ref{eqn:mbh.final}) assuming a simple relation between relic BH mass and peak quasi-star mass ($M_{\rm BH} = 0.1\,\epsilon_{0.1}\,M_{\rm q}$); (5) Claimed IMBH ``detection'' masses or upper limits. References labeled ``no evidence'' argue there is no positive evidence for an IMBH but set weak upper limits (shown). In all cases our predicted relic $M_{\rm BH}$ is below present upper limits (and claimed detections). References:
    [1] \protect\cite{zocchi_radial_2017}; 
    [2] \protect\cite{lutzgendorf_re-evaluation_2015}; 
    [3] \protect\cite{henault-brunet_black_2019}; 
    [4] \protect\cite{baumgardt_dynamical_2003}; 
    [5] \protect\cite{gieles_mass_2018}; 
    [6] \protect\cite{sollima_global_2017}; 
    [7] \protect\cite{mann_multi-mass_2019}; 
    [8] \protect\cite{baumgardt_n-body_2017}; 
    [9] \protect\cite{marel_new_2010}; 
    [10] \protect\cite{lanzoni_velocity_2013}; 
    [11] \protect\cite{cseh_radio_2010}; 
    [12] \protect\cite{bozzo_igr_2011}; 
    [13] \protect\cite{kiziltan_intermediate-mass_2017}; 
    [14] \protect\cite{gebhardt_intermediate-mass_2005};
    [15] \protect\cite{kamann_muse_2016}; 
    [16] \protect\cite{kirsten_no_2012}.
    [17] \protect\cite{2017MNRAS.468.2114P}.
    [18] \protect\cite{2013A&A...552A..49L}.
    }
    \label{tab:predictions}
\end{table*}

\subsection{Analytic Scalings}

Assuming the ``universal'' profile parameters, the contours of constant $M_{q}$ are approximately power-laws over most of the dynamic range of interest. We can approximate this quite well via a simple purely-analytic estimate for $M_{q}$ if we assume a single-power law profile with $\eta_1=2$ (isothermal), neglect ``edge'' effects (assume the stars outside $r_{\rm h}$ are not sinking efficiently), and approximate the effects of various non-linear terms like varying coulomb logarithms, the finite N-body point mass correction, finite IMF sampling, finite size of the quasi-star, and others as a systematic factor of $\sim 2$ normalization correction (reasonably well-motivated by our comparison in Fig.~\ref{fig:mcut}). With all of these approximations, we obtain the very simple expression:
\begin{align}
\label{eqn:mbh.final} M_{\rm BH} &= 0.1\,\epsilon_{0.1}\,M_{q} \\
\nonumber &\sim 250\,M_{\odot} \,\epsilon_{0.1} \left( \frac{M_{\rm cl,\,5}}{r_{\rm h,\,pc}}\right)^{3/4} \sim 250\,M_{\odot}\,\epsilon_{0.1}\,\left( \frac{V_{\rm eff}}{20\,{\rm km\,s^{-1}}} \right)^{3/2}
\end{align}
where $V_{\rm eff}^{2} \equiv G\,M_{\rm cl}/r_{\rm h}$, $M_{\rm cl,\,5}=M_{\rm cl}/10^{5}\,M_{\odot}$, $r_{\rm h,\,pc}=r_{\rm h}/{\rm pc}$.

Despite the many simplifications involved in deriving this expression, it provides a quite reasonable order-of-magnitude approximation to the most important results from our more detailed full model calculations.

\subsection{Globulars and ``Typical'' Dense Star Clusters}

Fig.~\ref{fig:sweetspot} \&\ Eq.~\ref{eqn:mbh.final} do suggest IMBHs could form in massive GCs, with typical masses $M_{\rm BH} \sim 0.0003\,\epsilon_{0.1}\,M_{\rm cl}$. In Fig.~\ref{fig:sweetspot} we also show the criterion $t_{\rm rlx} < 100\,$Myr which \citet{portegies_zwart_runaway_2002} and \cite{gurkan_formation_2004} argue is required for a GC with an initially {\em flat} ($\eta_1 = 0$), Plummer-like density profile to undergo any significant runaway merging. We see, as noted in \S~\ref{sec:intro}, that almost no known present-day massive clusters meet this criterion. The reason our modeling here predicts they {\em can} form central objects is because we argue they likely had steeper slopes at initial formation (allowing some merging near their center at these early times). But, essentially by definition, any interior region which has a steep enough slope to produce runaway merging within $<3\,$Myr will, on timescales $\sim $\,Gyr, have undergone relaxation, flattening the central profile seen today. 

However, although our models predict runaway merging {\em could} occur in the centers of almost all clusters at formation, the actual mass which we predict successfully merges (for realistic cluster $M_{\rm cl}$ and $r_{\rm h}$) is quite modest, giving a rather low $M_{\rm BH}/M_{\rm cl}$ compared to the clusters which undergo ``complete runaway core collapse'' (with $t_{\rm rlx} < 25\,$Myr) in \citet{portegies_zwart_runaway_2002}. In Table~\ref{tab:predictions}, we explicitly list a number of individual observed GCs from Fig.~\ref{fig:sweetspot} which have claimed detections or upper limits for central IMBHs. Many contradictory observational claims exist for some clusters, a well-known issue in the literature. Using the same models from Fig.~\ref{fig:sweetspot}, and the observed cluster properties, we give our best estimate of $M_{\rm BH}$ (approximately given by Eq.~\ref{eqn:mbh.final}), and compare to these observations. We see that the predicted relic mass from our calculations is typically $\sim 10^{-4}-10^{-3}\,M_{\rm cl}$, in many cases a factor $\sim 10$ or more below the claimed detections/upper limits. There is no case where our predicted $M_{\rm BH}$ exceeds even the most stringent upper limits. The most constraining examples we find are M15 and NGC 6388: here our ``default'' prediction is only a factor $\sim 2.5$ below the current upper limits or smallest values among the claimed detections. This implies $\epsilon_{0.1} \lesssim 3$ (i.e.\ $M_{\rm BH} \lesssim 0.3\,M_{q}$, if our models are to be believed at this level of accuracy). 

If some of the most-massive detections (with claimed IMBH masses up to $\sim 70$ times larger than our prediction) are indeed correct (although almost all of these cases are controversial with much lower limits claimed by other studies), then it would most likely imply that the central BHs grew rapidly after formation via some other process such as gas accretion (from e.g.\ stellar mass-loss in the cluster).

\subsection{Connection to SMBHs and More Massive Stellar Systems}

The predicted quasi-stars/IMBHs suggested in Fig.~\ref{fig:sweetspot} \&\ Eq.~\ref{eqn:mbh.final} become more massive, on average, in more massive systems, reaching $M_{\rm q} \sim 3\times10^{4}\,M_{\odot}$ in the most massive and dense NSCs and UCDs, and up to $M_{q} \sim 3\times10^{5}\,M_{\odot}$ in the centers of the most compact local and high-redshift bulges/Es. 

These are systems which are known to host super-massive BHs, with $M_{\rm BH} \sim 10^{4}-10^{10}\,M_{\odot}$ unambiguously detected (with the smallest BHs in dwarf NSCs, the most massive in compact Es), obeying a tight correlation with the velocity dispersion $\sigma$ of the surrounding stars $M_{\rm BH}^{M-\sigma} \sim 3\times10^{8}\,M_{\odot}\,(\sigma / 200\,{\rm km\,s^{-1}})^{4.3}$ \citep{kormendy_coevolution_2013}. However, noting that $\sigma \approx V_{\rm eff}$ in Eq.~\ref{eqn:mbh.final} for an isothermal profile, this implies that the present-day SMBHs observed are much more massive than the IMBH we predict from runaway merging at formation, for any $\sigma \gtrsim 7\,\epsilon_{0.1}^{0.3}\,{\rm km\,s^{-1}}$ (present-day $M_{\rm BH}^{M-\sigma} \gtrsim 100\,\epsilon_{0.1}^{1.5}\,M_{\odot}$). In other words, while the masses here are potentially interesting for very first initial seeds of the SMBHs, runaway merging {\em cannot} establish most of the mass of {\em any} observed BHs on the BH-host galaxy (or BH-NSC) scaling relations. It does not substantially reduce the amount of BH accretion (nor the time required for that accretion, if it occurs at e.g.\ a fixed Eddington ratio), nor even radically change the initial seed mass relative to commonly-assumed $M_{\rm BH}^{\rm seed} \sim 100\,M_{\odot}$ as the ``most optimal normal stellar relic'' remnant mass. 

An important additional caveat in these massive systems is that our models assume the stars are approximately co-eval. This is reasonable in the centers of dense GCs where the dynamical times are $\ll$\,Myr. However, in e.g.\ elliptical galaxy centers, the dynamical times can be $\gg 3\,$Myr; since stars cannot form much faster than the dynamical time, it is almost certainly the case that the massive star formation was extended in time relative to the lifetime $\sim 3\,$Myr of the quasi-star. In the center, later-forming stars can still sink, but they will merge with a central IMBH instead of quasi-star, producing a tidal disruption event and building up an accretion disk rather than directly forming a quasi-star.

\section{Conclusions}
\label{sec:conclusions}

Using the outputs of high-resolution numerical hydrodynamic simulations of star cluster/complex formation and destruction which have been shown to reproduce a wide range of GMC and cluster observables, we develop a semi-analytic model for the sinking of massive stars to cluster centers and their merger into a massive quasi-star. We find:
\begin{enumerate}
	\item The mass profile of YMCs ``at formation'' (centered on local peaks, as there can still be substructure) can be described by a double power-law with steep, near-isothermal inner slopes common (which flatten at later times as the inner regions dynamically relax). This means that some runaway merging can occur early even in clusters with relatively low mean densities and long relaxation times (e.g.\ $M_{\rm cl} \sim 10^{5}-10^{6}\,M_{\odot}$, $r_{\rm h} \sim 1-10\,$pc, with relaxation times $t_{\rm rlx} \gtrsim 100\,$Myr). The runaway ceases at $\sim 1-3\,$Myr, regardless of the details of the quasi-star evolution.
	\item Over the parameter space of greatest interest (where massive, dense stellar systems are observed), our predictions can be approximated with a simple scaling, with the total mass of massive stars which sink to the center and could potentially merge, $M_{q}$, scaling as $M_{q} \propto V_{\rm eff}^{1.5}$, where $V_{\rm eff}$ is a characteristic circular velocity (Eq.~\ref{eqn:mbh.final}). 
	\item Although some runaway merging is predicted in nearly all clusters (Fig.~\ref{fig:sweetspot}), the actual masses of IMBH relics predicted in our model for observed globulars and typical dense star clusters are quite modest, $\sim 100-1000\,M_{\odot}$. For relic mass $M_{\rm BH} \lesssim 0.3\,M_{q}$ (expected allowing for mass-loss, imperfect merging, quasi-star evolution, etc.), our predictions are consistent with even the most stringent upper limits (to our knowledge) on central IMBH mass in all clusters for which such constraints exist. The most constraining clusters for our models at present are M15 and NGC6388; the only well-studied cluster where our model predicts $M_{\rm BH} \gtrsim 1000\,M_{\odot}$ is G1. 
	\item In more massive systems such as nuclear star clusters, ultra-compact dwarfs, and the centers of compact ellipticals, the central object mass could reach $M_{q} \sim 10^{4}-10^{5}\,M_{\odot}$, an interesting range for initial seeds of super-massive BHs. However for {\em any} system with velocity dispersion $\gtrsim 10\,{\rm km\,s^{-1}}$, the SMBHs on the various observed SMBH-host scaling relations (e.g.\ $M_{\rm BH}-\sigma$) are far more massive than even the most optimistic IMBH masses resulting from runaway merging. Thus runaway merging does not significantly reduce the need for subsequent accretion to super-massive BH masses.
\end{enumerate}

Our models are intentionally simplified in order to survey a wide parameter space efficiently and guide intuition and predictions for future models. Using the models here to identify the most interesting parameter space, in future work we hope to consider explicit N-body simulations of the merging process in cluster centers (necessarily limited to a small number of realizations). In the systems where merging occurs most rapidly, it is also possible that mergers occur even {\em as} stars are still forming in the cluster, potentially before massive protostellar cores even complete their pre-main sequence evolution. Exploring this will require fully hydrodynamic+N-body simulations of star formation which can resolve the stellar IMF self-consistently and follow mergers as they occur ``live.'' Considerable uncertainties also still surround the actual dynamics of massive stellar mergers (including complicated effects not followed here, such as the effect of resolved binaries and hierarchical multiples on merger efficiency) and the evolution (especially as it grows via rapid merging) of the (proto) quasi-star. In addition, if such a system forms, a variety of processes may allow for rapid growth even after it collapses to an IMBH, as it could accrete tidally-disrupted lower-mass stars which sink on longer timescales (e.g.\ $m\sim 2-8\,M_{\odot}$), or stellar mass-loss products from AGB stars that can remain gravitationally bound in the cluster potential. All of these remain important subjects for future study.

\acknowledgments{Support for YS \&\ PFH was provided by NSF Collaborative Research Grants 1715847 \&\ 1911233, NSF CAREER grant 1455342, NASA grants 80NSSC18K0562, JPL 1589742. Numerical calculations were run on the Caltech compute cluster ``Wheeler,'' allocations FTA-Hopkins supported by the NSF and TACC, and NASA HEC SMD-16-7592.}

\bibliography{ms}

\end{document}